\documentclass{iopart}
\usepackage{cite}
\newcommand{\pdif}[2]{\frac{\partial #1}{\partial #2 } }
\newcommand{\eq}{\mbox{\scriptsize eq}}
\newcommand{\kb}{k_{\mbox{\scriptsize B}}}
\newcommand{\dd}{{\mbox{d}}}
\newcommand{\lgle}{\left\langle}
\newcommand{\rgle}{\right\rangle}
\newcommand{\eff}{\mbox{\scriptsize eff}}

\begin{document}

\title[]{Brownian motion with multiplicative noises revisited}

\author{T Kuroiwa$^1$ and K Miyazaki$^2$}

\address{$^1$ Institute of Physics, University of Tsukuba, Tennodai
1-1-1, Tsukuba 305-8571, Japan \\
$^2$ Department of Physics, Nagoya University, Furo-cho, Nagoya 464-8602, Japan}
\ead{$^1$ \mailto{s1030084@u.tsukuba.ac.jp}, $^2$ \mailto{miyazaki@r.phys.nagoya-u.ac.jp}}

\begin{abstract}
The Langevin equation with multiplicative noise and state-dependent transport coefficient
has to be always complemented with the proper interpretation rule of the
 noise, such
as the It\^o and Stratonovich conventions.
Although the mathematical relationship between the different rules 
and how to translate from one rule to another are well-established, 
it still remains controversial what is a more {\it physically} natural rule.
In this communication, we derive the overdamped Langevin equation with
 multiplicative noise for Brownian particles, by
systematically eliminating the fast degrees of freedom of 
the underdamped Langevin equation.
The Langevin equations obtained here vary depending on the choice of the noise
conventions but they are different representations for an identical phenomenon. 
The results apply to multi-variable, nonequilibrium, non-stationary systems, and other general settings.
\end{abstract}
\submitto{\JPA}

When the noise in a stochastic differential equation  of a
Markovian process is coupled nonlinearly with stochastic variables, it
is called multiplicative. 
A multiplicative Langevin equation is written in a general form of 
\begin{equation}
\dot{x}= f(x) + g(x)\xi(t),
\label{eq.1.1}
\end{equation}
where $f(x)$ is a systematic term, $\xi(t)$ is a white and Gaussian noise characterized by 
\begin{equation}
\lgle \xi(t)\rgle =0
\hspace*{0.5cm}{\mbox{and}}\hspace*{0.5cm}
\lgle \xi(t)\xi(t')\rgle= \delta(t-t'), 
\label{eq.1.white}
\end{equation}
and $g(x)$ is an $x$-dependent amplitude of the noise.
The multiplicative processes are ubiquitous in nature; 
nonlinear chemical reactions~\cite{Horsthemke1984}, diffusion processes 
with hydrodynamic interactions~\cite{Ermak1978jcp}, with geometric
constraints\cite{Morse2004,Namiki1984,Ciccotti2005,Polettini2013a}, in inhomogeneous
and nonequilibrium environments~\cite{Lancon2001,Volpe2010,Matsuo2000,Kampen1988},
to name but a few.

Equation (\ref{eq.1.1}) as it stands is meaningless unless one specifies the
interpretation of the multiplicative noise~\cite{Kampen2007}. 
More rigorously, (\ref{eq.1.1}) should be written using the
stochastic integration as
\begin{equation}
\Delta x = 
f(x)\Delta t + g(x^{\ast})\Delta W(t),
\label{eq.1.2}
\end{equation}
where $\Delta x =x(t+\Delta t)- x(t)$. $\Delta t$ is the increment of time
which is assumed to be sufficiently small. 
$\Delta W(t)$ is the increment of a Wiener process given by
\begin{equation}
\Delta W(t) = \int_{t}^{t+\Delta t}\!\!\dd s~ \xi(s),
\label{eq.1.wiener}
\end{equation} 
which satisfies $\Delta W^2= \Delta t$ (in a mean-square sense)~\cite{Kampen2007,Gardiner2009}. 
$x^{\ast}$ in $g(x^{\ast})$ is a midpoint value between
$x(t)$ and $x(t+\Delta t)$ defined by  
\begin{equation}
x^{\ast} \equiv \alpha x(t+\Delta t) + (1-\alpha) x(t).
\label{eq.1.xstar}
\end{equation}
One has to decide which value of $\alpha$ should be employed
in order to have the Langevin equation well-defined. 
A different value of $x^{\ast}$ leads to a different solution since 
$\Delta W$ is of the order of $\sqrt{\Delta t}$ and thus the correction
due to a choice of $x^{\ast}$ becomes the order of $\Delta t$.
Historically, three values of $\alpha$ have been most commonly adopted; 
the It\^o ($\alpha=0$), Stratonovich ($\alpha=1/2$), and the anti-It\^o
or isothermal ($\alpha=1$) conventions~\cite{Kampen1981,Hanngi1978,Lancon2001}. 
Although the mathematical relationship between the different choices of
$\alpha$ and how to map from one convention to others are
well-established~\cite{Gardiner2009}, it still remains controversial what value of
$\alpha$ is physically (not mathematically) more meaningful or favourable than others.
This controversy has been often called the It\^o-Stratonovich dilemma~\cite{Kampen2007,Kampen1981,Mannella2012,Lau2007,Kupferman2004,Suweis2011}.
The physically correct interpretation of the noise 
depends 
on the
physical problems~\cite{Kampen1981,Mannella2012,Mannella2011prl}. 
If the noise is of the external origin, its properties and
interpretation should be prescribed as a part of the modelling of the
system~\cite{Kampen1981,Kupferman2004,Sancho2000epjb,Carrillo2003pre}. 
The controversies appear only when 
calculations 
are either incorrect or
wrongly interpreted. 
On the other hand, when the noise is the internal one, {\it i.e.},
thermal fluctuations due to the coupling with a surrounding environment, 
the properties of the noise depend on the interplay of the noise 
with the energy dissipation
({\it e.g.}, through the fluctuation-dissipation theorem (FDT))
 and the interpretation of the noise is 
 subtle. 
This is the situation which we shall consider in this communication.
In the framework of the nonequilibrium thermodynamics, $f(x)$ is
often assumed to be given by a macroscopic law,  known as the celebrated
Onsager's regression hypothesis~\cite{Onsager1931a}. 
This 
hypothesis becomes less obvious when the noise is multiplicative.
This longstanding issue  is recently resurged as the new experimental
technique enables us to probe the effects of the nonlinear fluctuations
at mesoscopic scales and their consequences directly~\cite{Lancon2001,Volpe2010,Tupper2012}.
A canonical example is a Brownian motion of a colloidal particle
with the position-dependent friction coefficient.
The $x$-dependence can arise, for example, when the particle diffuses near a wall as the hydrodynamic interaction between the
particle and the wall makes the friction force sensitive to the distance
between them. 
Lau {\it et al.} have shown that the Langevin equation for the Brownian
particle for such situations is written as~\cite{Lau2007} 
\begin{equation}
  \dot{x} =- \Gamma(x)\pdif{U(x)}{x} + \kb T (1-\alpha)\pdif{\Gamma(x)}{x} +  g(x^{\ast})\xi(t), 
\label{eq.1.Lau}
\end{equation}
where $\Gamma(x)$ is (the inverse of) the position-dependent friction
coefficient, $U(x)$ is a potential of the system, $k_B$ is the
Boltzmann constant and $T$ is temperature of the system. 
$g(x)$ satisfies the FDT, $g^{2}(x) = 2\kb T \Gamma(x)$.
(\ref{eq.1.Lau}) is valid for any noise conventions, or
arbitrary $\alpha$. 
This expression can be derived using the condition that the
ensemble  average of the equation should vanish when the process is
stationary and the system is at thermally equilibrium.
It can be also reduced directly from the corresponding
Smoluchowski equation under the equilibrium condition~\cite{Gardiner2009}.
Note that the systematic term of the equation is not only given by the
macroscopic law (the first term of the right-hand side) but
also depends on the thermal fluctuation (the second term proportional to $\kb T$), unless $\alpha=1$.
As discussed above, any interpretation or any value of $\alpha$ is
mathematically admissible in (\ref{eq.1.Lau}).
However, Volpe {\it et al.} have argued that the anti-It\^o ($\alpha=1$)
convention should be taken from a physical point of view, 
since the systematic part of the Langevin equation should be governed solely by the macroscopic law~\cite{Volpe2010}.
Ermak {\it et al.} have derived (\ref{eq.1.Lau}) implicitly assuming
the It\^o convention ($\alpha=0$) in their classic paper~\cite{Ermak1978jcp}.
Others have claimed that the Stratonovich convention ($\alpha=1/2$) should be adopted~\cite{Hess1978,Sancho1982,Sancho2011}.  
They have ``derived'' this by starting from the underdamped Langevin equation for the coordinate and 
momentum and then adiabatically eliminating the latter in the overdamped limit where the friction is large.
One may argue that the choice of $\alpha$ is a matter of taste for the
equilibrium system.
However, the problem is less trivial and could be serious for nonequilibrium or other general situations,
since the form of the 
systematic term $f(x)$ in (\ref{eq.1.1}) is not known {\it a priori} and one has to derive
it from more microscopic equations or construct it by empirical modelling.

The goal of this communication is  to provide a general prescription to derive (\ref{eq.1.Lau}) 
and its generalized versions extended to nonequilibrium, multi-variable, and non-Cartesian coordinate systems. 
We also terminate controversies 
over the interpretation
of multiplicative noises 
by demonstrating that there is neither more physically meaningful nor natural choice of $\alpha$ 
in (\ref{eq.1.Lau}).
Our strategy is to adiabatically eliminate the momentum from the
underdamped Langevin equation, following the idea of 
~\cite{Ermak1978jcp,Hess1978,Sancho1982} 
but with a special care for the conventions of the noises.
We also point out a missing correction term in calculations reported
before~\cite{Hess1978,Sancho1982,Sancho2011}.   
Deriving the overdamped equation by integrating over the momentum of the
underdamped counterpart is hardly new. 
Commonly adopted method is to start from the Fokker-Planck equation to
obtain the Smoluchowski equation by integrating over the
momentum~\cite{Murphy1972,Titulaer1980,Matsuo2000},   
but it does not clarify the confusion of the noise interpretation.
The advantages of our method are that it is straightforward
and easily applicable for virtually all physical settings, aside  from
the equilibrium systems, since we bypass using the probability distribution function.
It also enables us to trace easily the origin of the specific convention of
the multiplicative noises.

We start with the most general form of the underdamped Langevin equation for generalized coordinates and momenta 
$(\vec{x}, \vec{p}) \equiv (x_1, \cdots, x_N,p_1,\cdots, p_N)$ with $N$ degree of freedom;
\begin{equation}
\left\{~
\eqalign{
\dot{x}_i = \pdif{H}{p_i}, 
\cr
\dot{p}_i = - \zeta_{ij}(\vec{x})\pdif{H}{p_j}  - \pdif{H}{x_i} + d_{ij}(\vec{x}) \xi_{j}(t).
}
\right.
\label{eq.1.full}
\end{equation}
Hereafter summation over repeated indices is adopted. 
$\xi_i(t)$ is a white Gaussian noise which satisfies $\langle
\xi_i(t)\rangle=0$ and
$\langle\xi_i(t)\xi_j(t')\rangle=\delta_{ij}\delta(t-t')$. 
$H=H(\vec{x},\vec{p})$ is a Hamiltonian of the system. 
The only prerequisite condition which we require is that 
 the kinetic part of the Hamiltonian has a
bilinear form, so that
\begin{equation}
 H(\vec{x}, \vec{p}) = \frac{1}{2}p_{i}m_{ij}^{-1}(\vec{x})p_j + U(\vec{x}).
\end{equation}
Note that the mass $m_{ij}(\vec{x})$ is a symmetric matrix and also
can be a function of the coordinate.
Such a coordinate dependence arises, for example, when the non-Cartesian
coordinate is used or the geometric constraints are present.
A typical example of the latter is the Brownian motion of the rigidly
bonded bodies, such as a string of a polymer chain~\cite{Morse2004,Kuroiwa2013b}.
$\zeta_{ij}(\vec{x})$ in (\ref{eq.1.full})
is a friction coefficient matrix.
If the system is at equilibrium, the FDT,
\begin{equation}
 d_{ik}(\vec{x})d_{kj}^{\dagger}(\vec{x})  = 2\kb T \zeta_{ij}(\vec{x}),
\end{equation}
should hold, where ``${\dagger}$'' represents the transpose of a matrix. 
Argument in the following, however, does not require the equilibrium condition.
In typical situations, the acceleration, $\dot{p}_i$, is negligibly
small unless one is interested in the very short time dynamics. 
The reason why we still keep it and start with the underdamped Langevin
equation, (\ref{eq.1.full}), is that the equation is free from the noise convention.
The correction due to different 
$\vec{x}^{\,\ast}$ 
in $d(\vec{x})$ (or  $\alpha$) is the
order of ${\cal O}(\Delta t^{3/2})$ or higher and (\ref{eq.1.full})
remains intact.
Therefore, the symbol ``$\ast$'' has been left out from the noise term in (\ref{eq.1.full}).

First, we illustrate how to derive (\ref{eq.1.Lau}) 
with arbitrary $\alpha$ directly by adiabatically eliminating the
momentum, if the system is at equilibrium.
Let us start with the simplest case where $N=1$ and the mass is constant. 
Equation (\ref{eq.1.full}) is written as 
\begin{equation}
\left\{~
\eqalign{
\dot{x} = p,
\cr
\dot{p} = - \zeta(x)p  - \pdif{U(x)}{x} + d(x) \xi(t),
}
\right.
\label{eq.1.full-1}
\end{equation}
where we set $m=1$. 
The goal is to derive the equation for $\Delta x(t) = x(t+\Delta t)-x(t)$ 
up to the linear order in $\Delta t$ in the overdamped limit. 
The overdamped limit means  that the time scale of the damping of the
momentum is much faster than the time scale we consider, {\it i.e.}, 
the limit of $\zeta \Delta t  \rightarrow \infty$ should be taken before
$\Delta t \rightarrow 0$.  
If the noise is additive and $d$ is independent of $x$, the
overdamped equation can be obtained by simply setting $\dot{p}=0$ in (\ref{eq.1.full-1}).
For the multiplicative case, it does not work. 
We have to choose a reference point $x^{\ast}$ defined by (\ref{eq.1.xstar}),
around which the correction should be carefully assessed.
Temporarily, we consider the case of $\alpha=1$, or $x^{\ast}=x(t+\Delta
t)$, to simplify the calculation.
This is also a choice which has been adopted implicitly in \cite{Hess1978,Sancho1982}.
We expand $\zeta(x)$ and $d(x)$ in (\ref{eq.1.full-1}) around $x^{\ast}$
as
\begin{equation}
 \zeta(x(t)) = \zeta^{\ast} {-} \pdif{\zeta^{\ast}}{x} \Delta x^{\ast}(t) 
\hspace*{0.5cm}{\mbox{and}}\hspace*{0.5cm}
 d(x(t)) = d^{\ast} {-} \pdif{d^{\ast}}{x} \Delta x^{\ast}(t), 
\end{equation}
where $\Delta x^{\ast}(t)\equiv x^{\ast}- x(t)$.
The quantities with ``$\ast$'' are functions of $x^{\ast}$.
Integration of (\ref{eq.1.full-1}) over $t$ twice gives a formal solution for $\Delta x(t)$;
\begin{eqnarray}
 \Delta {x} (t)
= 
\int_{t}^{t+\Delta t}\!\!\dd t_1\int_{-\infty}^{t_1}\!\!\dd t_2~ G(t_1-t_2) 
\left\{ 
F(x(t_2)) 
+ d^{\ast}\xi(t_2)
\right.
 \nonumber\\
\left.
\hspace*{3.6cm}
+ \pdif{\zeta^{\ast}}{x} \Delta x^{\ast}(t_2) p(t_2)
{-} \pdif{d^{\ast}}{x} \Delta x^{\ast}(t_2) \xi(t_2)
\right\}, 
\label{eq.1.full-3}
\end{eqnarray} 
where $F(x)=-\partial {U(x)}/\partial{x}$ is the force and 
$G(t) \equiv \exp\left[ -\zeta^{\ast}t\right]$ 
is a propagator.
Note that in \cite{Hess1978,Sancho1982}, the third term in $\{\cdots\}$ 
on the right hand side of (\ref{eq.1.full-3}) has been missing.
If both of the third and fourth terms in (\ref{eq.1.full-3}) 
are absent, it reduces to an overdamped equation;
\begin{equation}
\eqalign{
 \Delta {x}(t)
= 
\frac{1}{\zeta^{\ast}}F(x^\ast)\Delta t + \frac{d^\ast}{\zeta^\ast}\Delta W(t),
}
\label{eq.1.full-4}
\end{equation}
where $\Delta W(t)$ is the increment of the Wiener processes defined by (\ref{eq.1.wiener}).
The two correction terms in (\ref{eq.1.full-3}), however, can not be neglected since
the contributions of $\Delta x$ and $p$ are of the order of $\sqrt{\Delta t}$.  
Substituting their lowest order solutions 
\begin{equation}
\fl
\hspace*{0.2cm}
\eqalign{
p(t)\!=\!\!\int_{-\infty}^{t}\!\!\!\!\dd t_1~ G(t-t_1)d^\ast\xi(t_1)
\hspace*{0.3cm}\mbox{and}\hspace*{0.3cm}
\Delta x^\ast(t)\!=\!\!\int_{t_2}^{t+\Delta t}\!\!\!\!\dd
 t_1\int_{-\infty}^{t_1}\!\!\!\!\dd t_2~ G(t_1-t_2)d^\ast\xi(t_2)
}
\label{eq.1.full-p}
\end{equation}
to the third term in the right hand side of (\ref{eq.1.full-3}), one obtains
\begin{eqnarray}
\int_{t}^{t+\Delta t}\!\!\dd t_1\int_{-\infty}^{t_1}\!\!\dd t_2~ G(t_1-t_2) 
\pdif{\zeta^{\ast}}{x} \Delta x^{\ast}(t_2) p(t_2)
=
\frac{d^{\ast 2}}{2\zeta^{\ast 2}}\pdif{\zeta^{\ast}}{x}
\Delta t.
\label{eq.1.full-correction1}
\end{eqnarray} 
Likewise, the fourth term is written as
\begin{eqnarray}
\hspace*{-0.4cm}
{-}\int_{t}^{t+\Delta t}\!\!\dd t_1\int_{-\infty}^{t_1}\!\!\dd t_2~ G(t_1-t_2) 
\pdif{d^{\ast}}{x} \Delta x^{\ast}(t_2) \xi(t_2)
=
- \frac{d^\ast}{\zeta^{\ast 2}}\pdif{d^{\ast}}{x}\Delta t.
\label{eq.1.full-correction2}
\end{eqnarray} 
Applying the FDT,  $d^2(x) =2\kb T
\zeta(x)$, 
(\ref{eq.1.full-correction1}) and (\ref{eq.1.full-correction2}) can be rewritten as
\begin{eqnarray}
\fl
\hspace*{1.0cm}
\frac{d^{\ast 2}}{2\zeta^{\ast 2}}\pdif{\zeta^{\ast}}{x}\Delta t
=- \kb T \pdif{\zeta^{\ast -1}}{x}\Delta t
\hspace*{0.5cm}{\mbox{and}}\hspace*{0.5cm}
- \frac{d^\ast}{\zeta^{\ast 2}}\pdif{d^{\ast}}{x}\Delta t
= \kb T \pdif{\zeta^{\ast -1}}{x}\Delta t,
\label{eq.1.full-correction-all}
\end{eqnarray} 
respectively. 
These two terms are identical aside from their signs, cancel each other, and therefore (\ref{eq.1.full-3}) becomes
\begin{eqnarray}
 \Delta {x} 
= 
-\Gamma(x)\pdif{U(x)}{x}\Delta t + g(x^\ast)\Delta W(t)
\label{eq.1.full-7}
\end{eqnarray}
or simply
\begin{equation}
 \dot{x} =-\Gamma(x)\pdif{U(x)}{x} + g(x^\ast)\xi(t),
\label{eq.1.full-antiItofinal}
\end{equation}
where we define
$\Gamma(x)\equiv 1/\zeta(x)$ and $g(x)\equiv \Gamma(x)d(x)$.
Note that the overdamped version of the FDT $g^{2}(x) = 2\kb T\Gamma(x)$ holds.
Equation (\ref{eq.1.full-antiItofinal}) is identical with (\ref{eq.1.Lau}) for the anti-It\^o convention ($\alpha=1$).
In \cite{Hess1978,Sancho1982,Sancho2011}, one of the two
correction terms, (\ref{eq.1.full-correction1}), was missing and only
(\ref{eq.1.full-correction2}) was taken into account.
The crucial step of derivation of (\ref{eq.1.full-antiItofinal}) was to pick up the reference
point $x^\ast = x(t+ \Delta t)$ and expand $x(t_2)$ around it in (\ref{eq.1.full-3}).
If one expands $x(t_2)$ around a different $x^\ast$ ($\alpha\neq 1$), one obtains a different result.
Derivation for arbitrary $x^{\ast}$ can be done almost the same way as above. 
Only difference is that one has to divide the correction $\Delta x^{\ast}(t_2)=x^{\ast}-x(t_2)$ to two contributions
as 
$\Delta x^{\ast}(t_2)
= \alpha\left\{ x(t+\Delta t)-x(t_2) \right\} 
- (1-\alpha)
\left\{ x(t_2)-x(t) \right\}$ and then calculate the contributions separately.
One has to be careful about the order and the range of the time integrations.
After some manipulation,
one finds that, for arbitrary $\alpha$, (\ref{eq.1.full-correction1}) is replaced by 
\begin{equation}
 -(1-2\alpha)\frac{d^{\ast 2}}{2\zeta^{\ast 2}}\pdif{\zeta^{\ast}}{x}\Delta t
=(1-2\alpha) \kb T \pdif{\zeta^{\ast -1}}{x}\Delta t
\label{160644_27Aug13}
\end{equation}
and (\ref{eq.1.full-correction2}) is replaced by 
\begin{equation}
 -\alpha\frac{d^\ast}{\zeta^{\ast 2}}\pdif{d^{\ast}}{x}\Delta t
= \alpha k_BT\frac{\partial \zeta^{\ast -1}}{\partial x}\Delta t.\label{160649_27Aug13}
\end{equation}
The final expression is
\begin{equation}
 \dot{x} =  {-}\Gamma(x) \pdif{U}{x} {+} \kb T(1-\alpha) \pdif{\Gamma(x)}{x} + g(x^{\ast})\xi(t),
\label{eq.1.full-arbtrary}
\end{equation}
which is identical to (\ref{eq.1.Lau}).
The derivation above demonstrates  that the overdamped
Langevin equation is uniquely determined 
and there is no room for 
controversy on interpretation of
multiplicative noise. 
A different value of $\alpha$ is a difference in the 
representation for the identical physical process and there is no 
such thing as a more physically favourable representation.
A caveat is that, even though the systematic term of (\ref{eq.1.full-arbtrary})
depends on the value of $\alpha$, its average  does not. 
Indeed, the ensemble average of (\ref{eq.1.full-arbtrary}) at short times
under a fixed initial condition, denoted as $\overline{\dot{x}}$, is given by
\begin{eqnarray}
 \overline{\dot{x}}
&=& {-}\overline{\Gamma(x)\pdif{U(x)}{x}}
{+} \kb T(1-\alpha) \overline{\pdif{\Gamma(x)}{x}}
 + \overline{g(x^{\ast})\xi(t)}
 \nonumber\\
&=&{-}\overline{\Gamma(x)\pdif{U(x)}{x}}
{+} \kb T \overline{\pdif{\Gamma(x)}{x}}.
\label{eq.1.drift}
\end{eqnarray}
This is derived by expanding $x^{\ast}$ in $g(x^{\ast})$ around $x(t)$.
The fact that there exists a correction due to the fluctuations, 
$\kb T\overline{\partial\Gamma(x)/\partial x}$, in second line of (\ref{eq.1.drift}) implies 
that the multiplicative processes in general does not comply with
the Onsager's regression hypothesis which states that the averaged decay of the
fluctuations obeys the macroscopic law~\cite{Onsager1931a}.
This correction term plays an essential role in deriving the Onsager's reciprocal
relation, as we shall see later.

We argued above that any value of $\alpha$ is equally qualified.
But one may be still tempted to chose $\alpha=1$ in 
(\ref{eq.1.full-arbtrary}) since the systematic term is expressed solely by 
the force and the fluctuation correction vanishes.
However, the absence of the fluctuation correction is accidental and it
is inevitable for the multi-variable system ($ N >1$).
It is straightforward to generalize the above calculation for $N > 1$ 
where the friction coefficient as well as the coefficient of the
multiplicative noise are tensor.
For arbitrary $\alpha$, we have
\begin{equation}
\fl
\hspace*{0.5cm}
 \dot{x}_i = - \Gamma_{ij}\pdif{U}{x_j} + \kb T  (1-\alpha)\pdif{\Gamma_{ij}}{x_j} 
  +\frac{\alpha}{2}\left(g_{ik}\pdif{g^\dagger_{kj}}{x_j}-\pdif{g_{ik}}{x_j}g^\dagger_{kj}\right)
  + g_{ij}(x^\ast)\xi_j(t),
\label{eq.1.general-N}
\end{equation}
where $\Gamma_{ij}(\vec{x})\equiv \zeta^{-1}_{ij}(\vec{x})$, $g_{ij}(\vec{x})=\Gamma_{ik}(\vec{x})d_{kj}(\vec{x})$,
and the FDT is given by $g\cdot g^{\dagger} = 2\kb T \Gamma$.
Note that the fluctuation correction does not vanish even for $\alpha=1$.
This expression also shows that the noise correction can not be written
in terms of $\Gamma_{ij}(\vec{x})$ or its derivative 
but it is a complicated combination of $g_{ij}(\vec{x})$.
However, the ensemble average of (\ref{eq.1.general-N}) for a fixed initial condition 
is simpler and independent of $\alpha$;
\begin{equation}
\overline{\dot{x}_i(t)}  = - \overline{\Gamma_{ij}(\vec{x})\pdif{U(\vec{x})}{x_j}} + \kb T
  \overline{\pdif{\Gamma_{ij}(\vec{x})}{x_j} }.
\label{eq.1.general-averageN}
\end{equation}
This expression again shows that the Onsager's
regression hypothesis is violated and the systematic term contains the
fluctuation correction.
The violation of the hypothesis is essential to the Onsager's reciprocal
relation, $\Gamma_{ij} = \Gamma_{ji}$. 
This can be readily demonstrated by following the proof of the reciprocal
relation in the original paper by Onsager~\cite{Onsager1931a}.

Up to here we have assumed that the system is at equilibrium and the
FDT holds. 
This condition is lifted easily.
If the system is out of equilibrium and the FDT does not hold, one can employ
(\ref{eq.1.full-correction1}) and (\ref{eq.1.full-correction2}) instead
of (\ref{eq.1.full-correction-all}).
The final result for $N=1$ is 
\begin{equation}
 \dot{x} = -\Gamma(x) \pdif{U}{x}
  +\frac{D(x)}{\Gamma(x)}\pdif{\Gamma(x)}{x}
-\alpha \pdif{D(x)}{x} + g(x^{\ast})\xi(t),
\label{eq.1.non-eq-N1}
\end{equation}
where we defined $D(x)\equiv g^2(x)/2$.
The result is equivalent with the expression recently reported by Yang
{\it et al.}~\cite{Yang2013}, where the correction terms in
(\ref{eq.1.non-eq-N1}) 
were
 derived using the stationary condition that
the averages of the underdamped Langevin equation should vanish.

The simplest example is the system under a temperature gradient for which
$D(x)= \kb T(x)/\zeta$~\cite{Sekimoto1999,Matsuo2000}. 
In this case, (\ref{eq.1.non-eq-N1}) is simply written as 
\begin{equation}
\dot{x} = 
 -\alpha \pdif{D(x)}{x} + \sqrt{2D(x)}\xi(t),
\label{eq.1.non-eq-N1-extra}
\end{equation}
if $U$ is absent.
This expression looks as if the temperature gradient does not cause the
drift of the particle 
since the average of (\ref{eq.1.non-eq-N1-extra}) leads to $\overline{\dot{x}}=0$.
But if we translate (\ref{eq.1.non-eq-N1-extra}) to the equation for the density field defined by
$\rho(r,t)=\lgle \delta(r - x(t))\rgle$ (or equivalently the
Fokker-Planck equation), one finds that 
\begin{equation}
 \pdif{\rho}{t}=\nabla ( D\nabla \rho +D_T \nabla T),
\label{eq.1.non-eq-N1-extra2}
\end{equation}
where $D_T=D\rho/T$.
This is the thermal diffusion equation for a dilute suspension with the
Soret coefficient given by 
$S_T \equiv D_T/\rho D=1/T$
~\cite{Dhont2004} 
and also equivalent with the
Fokker-Planck equation derived in \cite{vKampenIEEE}.  
Note that (\ref{eq.1.non-eq-N1-extra2}) is independent of $\alpha$ and the underlining phenomenon does
not depend on the noise interpretation.

Generalization to multi-variable case is straightforward and the final expression is 
\begin{equation}
\fl 
\hspace*{0.6cm}
\dot{x}_i=-\Gamma_{ij}(\vec{x})\pdif{U(\vec{x})}{x_j}+\pdif{\Gamma_{ij}(\vec{x})}{x_k}D_{kl}(\vec{x})\Gamma^{-1}_{lj}(\vec{x})
  -\alpha \pdif{g_{ij}(\vec{x})}{x_k}g^\dagger_{jk}(\vec{x})
  +g_{ij}(\vec{x}^{~\ast})\xi_j(t),
\label{eq.1.non-eq-N2}
\end{equation}
where $D(\vec{x}) \equiv g(\vec{x})\cdot g^{\dagger}(\vec{x})/2$ and we
have assumed that $\zeta^{-1}dd^\dagger$ is a symmetric matrix.

Finally, let us consider the case where the mass is $\vec{x}$-dependent. 
The system under a geometric constraint
and
 a non-Cartesian coordinate 
are typical examples~\cite{Morse2004,Namiki1984,Ciccotti2005,Polettini2013a}.
The underdamped Langevin equation for $N=1$, (\ref{eq.1.full}), is written as
\begin{equation}
\left\{~
\eqalign{
\dot{x} =  \frac{1}{m(x)} p,
\cr
\dot{p} = - \gamma(x)p- \pdif{U(x)}{x}
-\frac{1}{2}\pdif{m^{-1}(x)}{x}p^2+d(x)\xi(t), 
}
\right.
\label{eq.1.mass-dep}
\end{equation}
where $\gamma(x) \equiv \zeta(x)/m(x)$.
(\ref{eq.1.mass-dep}) is formally integrated over time as
\begin{eqnarray}
\fl
 \Delta {x} (t)
= 
\int_{t}^{t+\Delta t}\!\!\dd t_1
\frac{1}{m(x(t_1))}
\int_{-\infty}^{t_1}\!\!\dd t_2~ G(t_1-t_2) 
\left\{ 
F(x(t_2)) 
+\frac{m^{\prime}(x(t_2))}{2m^2(x(t_2))}p^2(t_2)
\right.
 \nonumber\\
\left.
\hspace*{2.0cm}
+ d^{\ast}\xi(t_2)
+ \gamma^{\ast\prime} \Delta x^{\ast}(t_2) p(t_2)
+ d^{\ast\prime}\Delta x^{\ast}(t_2) \xi(t_2)
\frac{}{}\right\}, 
\label{eq.1.full-mass-formal}
\end{eqnarray} 
where the primes denote differentiation with respect to $x$.
Note that several terms newly appear due to the $x$-dependence of $m(x)$.
Careful calculations of the corrections around the reference point $x^{\ast}$ reveal 
that $m(x(t_1))$ in the first integral is replaced by $m^{\ast}\equiv m(x^{\ast})$ and 
$\gamma^{\ast\prime}$ in the fourth term is replaced by
$\zeta^{\ast\prime}/{m^\ast}$.
This is since the correction of the former cancels exactly with that of
the latter. This cancellation holds for arbitrary $N$ and for the
nonequilibrium cases.
Finally the second term which is proportional to $p^2(t_2)$ can be computed
by substituting the first equation of (\ref{eq.1.full-p}) into
$p^2(t_2)$ and using
the fact
 that $\xi(t)$ is white Gaussian.
The result is nothing but the equipartition theorem 
$p^2(t) = {m(x(t))\kb T}$ (in a mean-square sense), if the system is at equilibrium.
These results simplify (\ref{eq.1.full-mass-formal}) substantially
and make the analysis essentially identical with that of the case of a
constant mass.
The final expression thus obtained is (\ref{eq.1.full-arbtrary}) but
with the potential term replaced by 
\begin{equation}
U(x) \longrightarrow  U_{\eff}(x) = U(x) -\frac{\kb T}{2} \log m(x).
\end{equation}
This shift of the potential energy is 
a natural consequence, since the canonical equilibrium distribution
functions should be $P_{\eq}(x) \propto \sqrt{m(x)}~\exp\left[-U(x)/\kb
T\right]=\exp\left[-U_{\eff}/k_{B}T\right]$.
An extension to $N > 1$ is obvious and $U(\vec{x})$ in
(\ref{eq.1.general-N}) should be replaced by 
\begin{equation}
 U_{\eff}(\vec{x}) = U(\vec{x}) -\frac{\kb T}{2} \log\det m(\vec{x}),
\end{equation}
where $\det m(\vec{x})$ is the determinant of the matrix $m_{ij}(\vec{x})$.

If the system is out of equilibrium, 
the equipartition theorem does not hold and the final expression becomes
far more complicated. 
For $N=1$, (\ref{eq.1.non-eq-N1}) should be replaced by
\begin{eqnarray}
\fl
\hspace*{0.4cm}
 \dot{x}&=& 
- \Gamma(x)\pdif{U(x)}{x}-\pdif{m(x)}{x}\frac{D(x)}{2m(x)}  
  +\frac{D(x)}{\Gamma(x)}\pdif{\Gamma(x)}{x}
  - \alpha \pdif{D(x)}{x}
  +g(x^\ast)\xi(t).
\end{eqnarray}
Likewise,  for $N>1$,
\begin{equation}
-\frac{1}{2}\Gamma_{ij}\pdif{m^{-1}_{kl}}{x_j}m_{km}D_{mn}\Gamma^{-1}_{nl}
\end{equation}
has to be added to the right-hand side of (\ref{eq.1.non-eq-N2}),
where we assume $m\zeta^{-1}dd^\dagger$ is a symmetric matrix. 

In this communication, we present a general prescription to derive the overdamped Langevin equation for 
Brownian particles with multiplicative noise by adiabatically 
eliminating the fast relaxing momentum. 
The method is  valid both in and out of equilibrium.
We found that there is no such thing as the It\^o-Stratonovich dilemma.
The Langevin equation varies depending on the choice of the noise
conventions, or $\alpha$, but they are just different representations for the identical phenomenon. 
There have been arguments, or at least misunderstanding, that $\alpha
=1$ (anti-It\^o) is more favourable from a physical point
of view since the systematic term is simply written in terms of the
macroscopic law. 
But this is not true as it is obvious for the multi-variable
systems, for which the systematic term can not be solely written by a
macroscopic law for any noise interpretation.
Note that the Onsager's regression hypothesis generally does not hold
for multiplicative processes. 
For the equilibrium systems, our result is not new, in the
hindsight, since the correct form of the Langevin equation can be prescribed from
the strong constraint that the corresponding probability distribution must
relax to the equilibrium distribution function~\cite{Mannella2012,Mannella2011prl}. 
The advantage of our method is that the derivation is far simpler and
more straightforward, and therefore, can be easily generalized to
nonequilibrium, non-stationary, and any other settings. 
Although we have considered the simple colloidal particle system, 
the formulation considered here can be applied to general class of
stochastic processes as long as the equations have a canonical structure
written in terms of the generalized momenta and coordinates. 
Thus the argument discussed here is of broad applicability.

\ack
We thank Yohei Fujitani for fruitful discussions and insightful comments. 
This work was supported by the JSPS Core-to-Core Program ``International
research network for non-equilibrium dynamics of soft matter''.
K. M. is supported by KAKENHI nos
24340098, 
25103005, 
and 25000002. 

\section*{References}
\providecommand{\newblock}{}

\end{document}